\documentstyle[aps,epsfig]{revtex}
\begin{document}
\draft
\author{B.~All\'es$^{\rm a}$, G.~Cella$^{\rm b}$, M.~Dilaver$^{\rm c}$,
Y.~G\"und\"u\c c$^{\rm c}$}

\address{$^{\rm a}$Dip.~Fisica, Sezione Teorica, 
 Universit\`a di Milano, Via Celoria 16, 20133 Milano, Italy}

\address{$^{\rm b}$Dip.~Fisica, Universit\`a di Pisa, Piazza Torricelli 2, 
 56126 Pisa, Italy}

\address{$^{\rm c}$Hacettepe University, Physics Department, 
 06532 Beytepe, Ankara, Turkey}
\title{Testing fixed points in the
	2D $O(3)$ non--linear $\sigma$--model.}
\maketitle
\begin{abstract}
Using high statistic numerical results we investigate the properties
of the $O(3)$ non--linear 2D $\sigma$-model. 
Our main concern is the detection of an hypothetical
Kosterlitz--Thouless--like ($KT$) phase transition which would contradict the
asymptotic freedom scenario. Our results
do not support such a $KT$--like phase transition.
\end{abstract}
	
\pacs{05.50.+q; 11.15.Ha; 12.38.Bx}

\section{Introduction}

The bidimensional $O(N)$ non--linear sigma model is defined on the lattice by
the action
\begin{equation}
\label{eq:S}
S = -\beta \sum_{x,\mu} \phi(x) \cdot \phi(x+\mu) \ .
\end{equation}
Here $\phi(x)$ is a $N$--component vector constrained by
$\phi^2(x)=1$. Perturbation theory predicts that this model is
asymptotically free if $N\ge3$~\cite{father1,father2}. 
Large distance observables must scale
accordingly with the renormalization group equation. In particular we
find for the correlation length
\begin{equation}
\label{eq:fxi}
\xi = C_\xi \left(\frac{N-2}{2\pi\beta}\right)^\frac{1}{N-2}
\exp\left( \frac{2\pi\beta}{N-2} \right) \left( 1 + \sum_{k=1}^{\infty} \frac{a_k}{\beta^k} \right)
\end{equation}
and for the magnetic susceptibility 
\begin{equation}
\label{eq:fchi}
\chi = C_\chi \left(\frac{N-2}{2\pi\beta}\right)^\frac{N+1}{N-2}
\exp\left( \frac{4\pi\beta}{N-2} \right) \left( 1 + \sum_{k=1}^{\infty} \frac{b_k}{\beta^k} \right)\ .
\end{equation}
The coefficients $a_k$ and $b_k$ can be evaluated perturbatively
starting from 3--loop order. They describe non--universal (scheme
dependent) corrections to asymptotic scaling. The constants 
$C_\xi$ and $C_\chi$
cannot be evaluated perturbatively. 
There are many possible definitions of correlation length. All of them
scale in the same way,~Eq.~(\ref{eq:fxi}) with the same coefficients $a_k$. The
constant $C_\xi$ is instead definition dependent. The exponential fall--off
of the correlation function at large distances provides the so--called
exponential definition whose constant
$C_{\xi}$ can be obtained exactly 
(alas not rigorously!) with the Bethe--Ansatz technique~\cite{hmn,hn}
\begin{equation}
\label{eq:cxi}
C_{\xi} = 
 \left(\frac{e}{8}\right)^\frac{1}{N-2}\Gamma\left( \frac{N-1}{N-2}\right)
 2^{-5/2} \exp\left(-\frac{\pi}{2(N-2)}\right) \ .
\end{equation}
$C_\chi$ is known at second order in the $1/N$
expansion~\cite{chinonp}.
In the ratio $\chi/\xi^2$ the exponential dependence on $\beta$ disappears, 
and the quantity
\begin{equation}
R_{PT} \equiv 
 \frac{\chi}{\xi^2} \left(\frac{2\pi\beta}{N-2}\right)^{\frac{N-1}{N-2}}
 \left( 1 + \sum_{k=1}^{\infty} \frac{c_k}{\beta^k} \right)
 \; {\buildrel {{\beta \to \infty}} \over \longrightarrow}
 \; \;\frac{C_\chi}{C_\xi^2}
\end{equation}
should stay constant for large enough $\beta$. The coefficients
$c_k$ can be easily calculated from $a_k$ and $b_k$.

According to the Mermin--Wagner theorem~\cite{merminw} the continuous
$O(N)$ simmetry cannot be broken in two dimensions.  Perturbative
expansions are constructed around a non--symmetric trivial vacuum, so
their validity is not guaranteed.  
By analysing the percolation properties of the Fortuin--Kasteleyn 
clusters~\cite{fk} it has been argued~\cite{seiler1,seiler2} that perturbative
methods are inadequate to describe the model. The main conclusion from these
analyses is that the non--linear sigma model in 2 dimensions has no mass--gap,
contradicting the results of~\cite{hmn,hn}, and that for any $N\geq 2$ 
it undergoes a $KT$--like phase transition at a finite inverse
temperature $\beta_{KT}$.

If this is the case, then the ratio
\begin{equation}
\label{eq:RKT}
 R_{KT} \equiv \frac{\chi}{\xi^{2-\eta}} (\beta_{KT}-\beta)^r
\end{equation}
should be asymptotically constant for $\beta_{KT}-\beta$
positive and small. Here $\eta$ is a
critical exponent. For the $O(2)$ model, where the $KT$ transition is
generally trusted, $\eta=1/4$. On the other hand, 
renormalization group considerations
and numerical evaluations of $r$ yield the 
bound $|r| \lesssim 0.1$ \cite{rgr1,rgr2,rn1,rn2,rn3}.

It is well known that in the $O(3)$ model there is not a good
numerical evidence of the onset of the asymptotic scaling
regime. In~\cite{psKT} the authors find that in this model the
numerical data for the ratio $R_{KT}$ with $\eta=1/4$ can be well fitted
with a constant, while the ratio $R_{PT}$ is clearly decreasing.
They make use of data obtained from a simulation with the standard
action~Eq.~(\ref{eq:S}).

In~\cite{ourwork} it has been shown however that $R_{KT}$ is not constant.
To do this, the 0--loop Symanzik improved action was numerically simulated
at high statistics ($O(10^7)$ measurements), including corrections to 
asymptotic scaling up to 3 loops. In~\cite{ourwork} we claimed that the
discrepancy on the constancy of $R_{KT}$ between~\cite{ourwork} 
and~\cite{psKT} originates from the higher resolution of the data 
in~\cite{ourwork}. Arguments against these conclusions has been raised 
based on possible distorting effects of the antiferromagnetic coupling
present in the improved action. Here we want to perform an analogous study,
at high statistics too, on the $O(3)$ model with standard action where 
such a coupling is absent.

 For completeness we will show the results for $C_\xi$, $C_\chi$ 
and the $R_{PT}$ ratio obtained from our data.

\section{Monte carlo simulation}

We have simulated the standard action~Eq.~(\ref{eq:S}) 
on lattice sizes $L$ by using the
Wolff algorithm~\cite{wolff1}. The wall--wall correlation function 
\begin{equation}
G(t) \equiv \frac{1}{L} \sum_{x}  \langle\phi(0,0)\cdot\phi(t,x)\rangle \ 
\end{equation}
has been measured with improved estimators~\cite{wolff2}.
We have always chosen $L$ big enough to avoid sizable finite volume
effects: $L/\xi\gtrsim 7$~~\cite{fsscp,ourwork}. 

Using these data we can evaluate the correlation function at momentum $p$
\begin{equation}
\label{eq:gtilde}
{\widetilde G}(p) \equiv L \sum_t e^{i p t} G(t) ,
\end{equation}
the magnetic susceptibility $\chi\equiv\widetilde G(0)$
and the correlation length from the exponential 
fall--off of~Eq.~(\ref{eq:gtilde})
\begin{equation}
 \xi \equiv -\lim_{t\rightarrow\infty} \frac{t}{\ln G(t)}.
\end{equation}
The data for these quantities are listed in
Table~\ref{tab:rawdata}, together with the energy density defined as 
\begin{equation}
\label{eq:E}
 E \equiv \langle \phi(0) \cdot \phi(0 + \hat \mu) \rangle
\end{equation}
where $\hat\mu$ is a generic direction and there is {\it no} a sum over that 
$\mu$.

\section{Results}

Although the present note is mainly concerned on the existence of a phase
transition at a finite temperature, we give for completeness also
the results for $C_\xi$, $C_\chi$ and $R_{PT}$.

\begin{figure}
\begin{center}
\leavevmode
\mbox{\epsfig{file=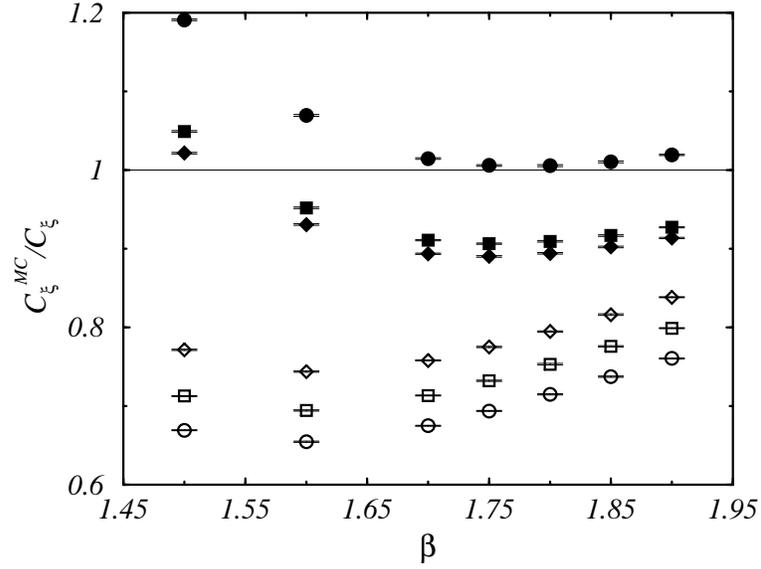,angle=270,width=0.65\textwidth}}
\end{center}
\caption{The ratio between non--perturbative 
constants $C_\xi^{MC}/C_\xi$ for the model. 
Empty circles (squares, diamonds) stand for 
the standard scheme 2--loop (3--loop,4--loop) results;
filled circles (squares, diamonds) stand for the 
energy scheme 2--loop (3--loop,4--loop) results.}
\label{fig:figXI}
\end{figure}

By using Eq.~(\ref{eq:fxi}) and our Monte Carlo data we obtain a
prediction about $C_\xi$, which we call $C_\xi^{MC}$. In
Fig.~\ref{fig:figXI} we plot with empty symbols the ratio
$C_\xi^{MC}/C_\xi$ versus $\beta$, adding scaling corrections up to
four loops~\cite{4loop} (we have taken into account the
corrections of~\cite{shin}). It is apparent that we are not in an
asymptotic scaling regime, although the non--universal scaling
corrections provide a systematic improvement. In the best (4--loop) case
the lack of asymptotic scaling is around $15\%$.

We have also used an effective scheme~\cite{effective}
defined by the perturbative expansion of the energy~Eq.~(\ref{eq:E})
\begin{eqnarray}
E &=&  1 - \frac{w_1}{\beta} - \frac{w_2}{\beta^2} - 
       \frac{w_3}{\beta^3}  - \frac{w_4}{\beta^4} - 
       O\left(\frac{1}{\beta^5}\right) , \\ \nonumber
\beta_E &\equiv& \frac{w_1}{1-E} = \beta - \frac{w_2}{w_1} + 
   \frac{w_2^2-w_1 w_3}{w_1^2} \frac{1}{\beta} -
   \frac{w_2^3-2 w_1 w_2 w_3+w_1^2 w_4}{w_1^3} \frac{1}{\beta^2} + 
   O\left(\frac{1}{\beta^3}\right) .
\end{eqnarray}
The results in this scheme are displayed as full symbols in Fig.~\ref{fig:figXI}.
Here the lack of asymptotic scaling is still large: about 7--8\% in the 
best case. 
 
\begin{figure}
\begin{center}
\leavevmode
\mbox{\epsfig{file=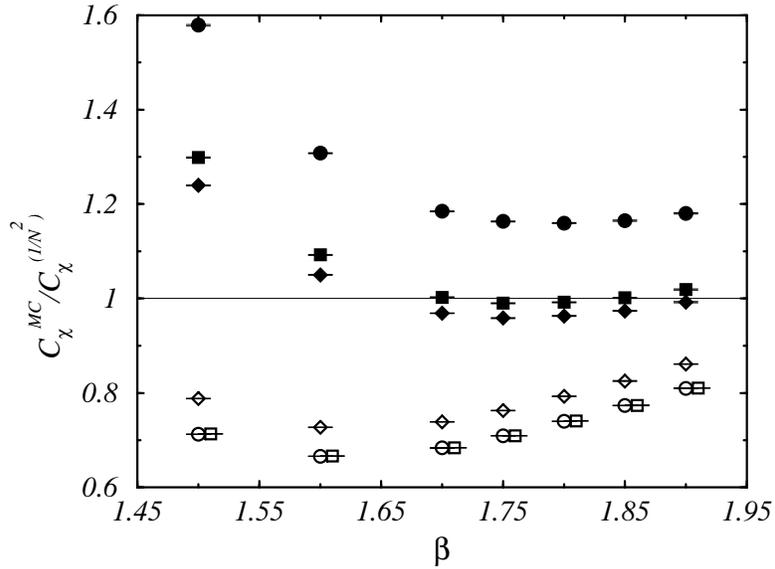,angle=270,width=0.65\textwidth}}
\end{center}
\caption{The ratio between non--perturbative constants 
$C_\chi^{MC}/C_\chi^{(1/N^2)}$ for the model. 
The meaning of the symbols is the same of Fig.~\ref{fig:figXI}.
Some data have been horizontally shifted to render the figure clearer.}
\label{fig:figCHI}
\end{figure}

Starting from Eq.~(\ref{eq:fchi}) we can study in the same way the ratio
$C_\chi^{MC}/C_\chi$, using for $C_\chi$ the $O(1/N^2)$ result~\cite{chinonp}
$C_\chi^{(1/N^2)} = 0.0127$.
The percentage of discrepancy in Fig.~\ref{fig:figCHI} 
between the Monte Carlo result and the ($1/N^2$) analytical
prediction is similar to that of the correlation length in 
Fig.~\ref{fig:figXI}.
It is interesting to note that the $3$--loop
correction is irrelevant, while the $4$--loop one is not.

The data in the effective energy--scheme 
for the Fig.~\ref{fig:figCHI} are displayed as full symbols.
Here we see an excellent agreement between the Monte Carlo data at 4--loops and
the $1/N^2$ approximation. 
We do not know how good is the $1/N^2$ approximation to $C_\chi$. Therefore
we avoid drawing optimistic conclusions about the approach to the
asymptotic scaling regime in this case.

\begin{figure}
\begin{center}
\leavevmode
\mbox{\epsfig{file=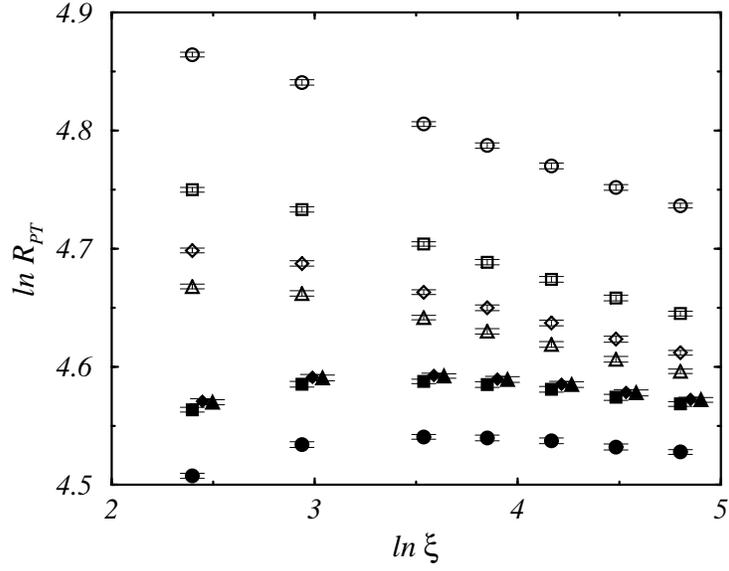,angle=270,width=0.65\textwidth}}
\end{center}
\caption{The $PT$ ratio for the model. Empty circles (squares, diamonds,
triangles) stand for the 1--loop (2--loop, 3-loop, 4--loop) results
in the standard scheme; filled circles (squares, diamonds,
triangles) stand for the 1--loop (2--loop, 3-loop, 4--loop) results
in the energy scheme.
Some data have been horizontally shifted to render the figure clearer.}
\label{fig:figRPT}
\end{figure}

In Fig.~\ref{fig:figRPT} we plot the logarithm of the ratio $R_{PT}$
versus the logarithm of the correlation length. 
Again the full (open) symbols correspond to the effective (standard) scheme.
We see that
data cannot be considered constant, although corrective terms give a
systematic improvement. We stress the fact that further
corrections and further statistics do not worsen the results.
 From the larger $\beta$ and considering the errors coming altogether 
from the 2--3--4 loop corrections, we get $\ln\left(C_\chi/{C_\xi}^2\right)=
4.5705(39)$ which, by using Eq.~(\ref{eq:cxi}), leads to the prediction 
$C_\chi=0.01507(6)$. This value however must be 
regarded as an upper bound because
data in Fig.~\ref{fig:figRPT} are still descending.

\begin{figure}
\begin{center}
\leavevmode
\mbox{\epsfig{file=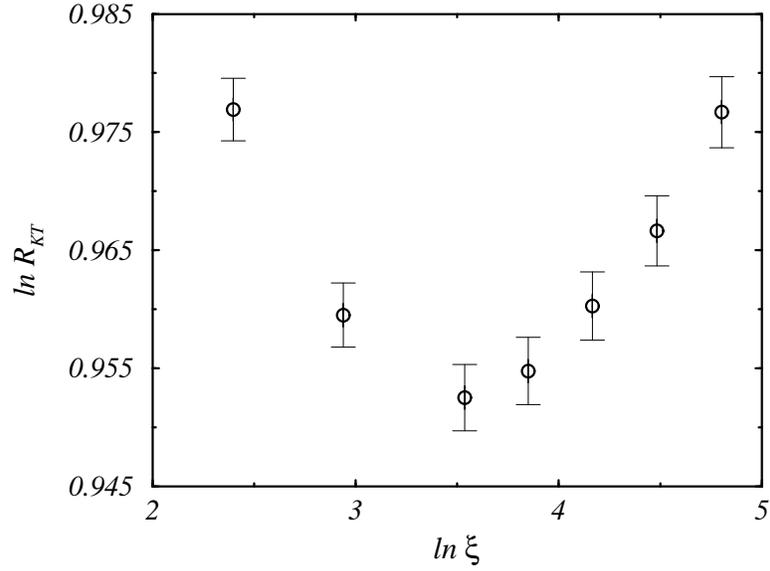,angle=270,width=0.65\textwidth}}
\end{center}
\caption{The $KT$ ratio for the model.}
\label{fig:figRKT}
\end{figure}

In Fig.~\ref{fig:figRKT} we plot the main result of the present paper: 
the ratio $R_{KT}$ versus $\xi$ in a
log--log scale. As it already happened for the Symanzik action, 
the shape of the curve indicates
a non-constant ratio, contradicting the claims of~\cite{psKT}. In calculating
this figure we have omitted the power $(\beta_{KT} -\beta)^r$, 
see~Eq.~(\ref{eq:RKT}).
Had the model to undergo a phase transition at finite coupling
$\beta_{KT}$, the correlation length $\xi$ should behave like
\begin{equation}
\label{eq:xiKT}
\xi = A \exp\left({B \over \sqrt{\beta_{KT} - \beta}} \right) .
\end{equation}
If we try to fit our data to such functional form we obtain very unstable
results (including more or less data points in the fit, the results for
$A$, $B$ and $\beta_{KT}$ vary strongly). The fit to the form 
Eq.~(\ref{eq:fxi}) is much more stable. This phenomenon has a simple
interpretation: $\beta_{KT}$ is so large that Eq.~(\ref{eq:xiKT}) can
be approximated to
\begin{equation}
\xi\approx A'\,\exp\left(B'\,\beta\right) \qquad \qquad
A'\equiv A \exp\left({B \over \sqrt{\beta_{KT}}}\right) \qquad \qquad
B'\equiv {B \over 2 \beta_{KT}^{3/2}}
\end{equation}
and therefore we are fitting actually the 
combination $B/\beta_{KT}^{3/2}$.
If $\beta_{KT}$ is so large and recalling that $|r|\lesssim 0.1$, 
then the power $(\beta_{KT}-\beta)^r$ is almost constant within the 
narrow interval of our working $\beta$'s. This is why we omitted such
a factor in the calculation of $R_{KT}$.

Notice that the shape of the curve for $R_{KT}$ can be obtained by using the
perturbative expressions in Eqs.~(\ref{eq:fxi},\ref{eq:fchi}). Owing
to the incomplete asymptotic scaling, we obtain a 20\%
difference in the overall scale in the best case (4 loops and energy
scheme). However the position of the minimum is well predicted: the 
perturbative expansion at 4 loops and energy scheme display a minimum
at $\ln\xi=3.2$ while the curve in Fig.~4 has minimum at 
$\ln\xi= 3.5(4)$ (the error has been estimated by interpolating
the top and bottom values of the error bars in the data of Fig.~4).

The prediction of \cite{seiler1,seiler2} is that the $O(N)$ model
behaves like the $O(2)$ model for $N\geq 3$. Therefore the set
of critical exponents of $O(2)$ should also apply for $O(3)$.
However one could imagine a set of slightly different exponents
(after all, the $O(3)$ and $O(2)$ models have different numbers 
of degrees of freedom). Following this idea, we have made further
fits on the data for $\xi$ to functional forms similar to Eq.~(\ref{eq:xiKT})
but with other powers of $(\beta_{KT}-\beta)$. In all cases the 
only sensible conclusion was that $\beta_{KT}$ is very large. One
can also try other versions of the ratio $R_{KT}$ by varying $\eta$.
The tendency is that for larger (smaller) $\eta$ the curve acquires
a positive (negative) slope for all points. We could not find a value
of $\eta$ for which the whole set of points in Fig.~\ref{fig:figRKT}
flattens out.

\section{Conclusions}

We have done a Monte Carlo simulation for the standard action of
the $O(3)$ non--linear
$\sigma$--model in 2-dimensions. We have improved the statistics with
respect to previous work taking advantage of the recently
calculated corrections to scaling~\cite{4loop,shin} and
energy~\cite{ourwork}.  We have tested the perturbation theory
predictions in both the standard and effective schemes. 

In Figs.~1--3 we have plotted
our results for $C_\xi$, $C_\chi$ and $R_{PT}$ 
as a function of $\beta$. We see that
the use of an effective scheme improves the asymptotic scaling in a sensible
way. However, in the best case, we are still far from asymptotic scaling
by roughly 10\%. Much closer approaches to the scaling region were obtained
by using the Symanzik action~\cite{ourwork}.

It is remarkable that in all cases, the 4--loop correction in the
effective scheme is negligible
and that if we trust the $O(1/N^2)$ estimate of $C_\chi$ then our
data have reached the scaling regime after including 3 or 4 loops.

As for our main result, Fig.~\ref{fig:figRKT}, it does not give any
support to the $KT$--scenario. The plot has the same shape than for the
Symanzik action~\cite{ourwork} and this indicates that distorting 
effects due to the antiferromagnetic coupling are quite unlikely.
In fact, in~\cite{ourwork} we worked with $\xi > 16$ in order to
avoid any strong coupling effect. 

As already shown in~\cite{ourwork}, the ratio $R_{PT}$ is not constant
but further corrections straighten it. This tendency is not satisfied
for the $R_{KT}$ ratio if we increase the statistics.
In conclusion, we think that
the main claim in~\cite{psKT} cannot be maintained within the present
status of numerical simulations because either $\beta_{KT}$ is too
large to see any effect of the $KT$--transition or even it is infinity.

\section{Acknowledgements}

B.A. wishes to thank the Theory Group of the Hacettepe University for their
kind hospitality.

\section{\it Note Added}

To obtain the results displayed in Figs.~1--3, we have used the corrected
values for the finite integrals published in~\cite{shin} needed for the
4--loop beta function. We notice that these corrected values do not change
appreciably the results shown in~\cite{ourwork}. For instance, for the
$O(8)$ model with standard action and energy scheme it was shown 
in~\cite{ourwork} that $C_\xi^{MC}$ differs from Eq.~(\ref{eq:cxi}) by
0.5\%; after the corrections of~\cite{shin} this difference becomes
0.4--0.5\%.

\newpage

\begin{table}
\caption{Results of Monte Carlo data for the $O(3)$ nonlinear $\sigma$--model
with standard action.}  
\label{tab:rawdata} 
\begin{tabular}{dddddd}
 \hline 
$\beta$ & $L$ & stat & $\chi$ & $\xi$ & $E$ \\ 
\hline
1.50 & 100 & $5\times10^6$    & 176.20(0.16) & 10.99(1) & 0.6015813(19) \\
1.60 & 150 & $5\times10^6$    & 446.91(0.42) & 18.89(2) & 0.6357033(11) \\
1.70 & 260 & $5\times10^6$    & 1264.4(1.00)  & 34.36(3) & 0.6642223(6)  \\
1.75 & 340 & $5\times10^6$    & 2189.5(2.2)& 46.97(5) &  0.6766299(4)  \\
1.80 & 450 & $5\times10^6$    & 3827.4(3.9)& 64.43(7) &  0.6879333(3)  \\
1.85 & 640 & $5\times10^6$    & 6717.6(7.1)& 88.53(10)&  0.6983241(2)  \\
1.90 & 860 & $5\times10^6$    & 11850.(13.) & 121.70(10) & 0.7079167(1)  \\
\hline
\end{tabular} 
\end{table}

\end{document}